\documentstyle[12pt]{article}
\def\Al{{\cal A}}
\def\Di{{\cal D}}
\def\Hi{{\cal H}}
\def\Ji{{\cal J}}
\def\Con{{\cal C}}
\def\Trip{{\cal T}}

\def\Id{{\bf Id}}
\def\im{{\bf i}}
\def\U{{\bf U}}
\def\V{{\bf V}}
\def\NCI{{\int \!\!\!\!\!-}}
\def\be{\begin{equation}}
\def\ee{\end{equation}}
\def\bea{\begin{eqnarray}}
\def\eea{\end{eqnarray}}
\def\beann{\begin{eqnarray*}}
\def\eeann{\end{eqnarray*}}
\begin{document}
%
\hfill {\bf UFRJ-IF-NCG-1999/1}  
\vskip 1cm
\begin{center} {\Large \bf On the Product of Real Spectral Triples}
\vskip 0,5cm {\bf \large F.J.Vanhecke}\footnote{\small E-mail: vanhecke@if.ufrj.br }
\\
\vskip 0.5cm
{\it  Instituto de F\'\i sica}\\
{\it UFRJ, Ilha do Fund\~ao, Rio de Janeiro, Brasil.}\\
\end{center}
\begin{abstract} 
The product of two real spectral triples $\{\Al_1,\Hi_1,\Di_1,\Ji_1,\gamma_1\}$ and $\{\Al_2,\Hi_2,\Di_2,\Ji_2(,\gamma_2)\}$, the first of which is necessarily even, was defined by
A.Connes \cite{Con2} as $\{\Al,\Hi,\Di,\Ji(,\gamma)\}$ given by $\Al=\Al_1\otimes\Al_2$, $\Hi=\Hi_1\otimes\Hi_2$, $\Di=\Di_1\otimes\Id_2+\gamma_1\otimes\Di_2$, $\Ji=\Ji_1\otimes\Ji_2$ and, in the even-even case, by $\gamma=\gamma_1\otimes\gamma_2$. Generically it is assumed that the real structure $\Ji$ obeys the relations ${\Ji}^2=\epsilon\Id$, $\Ji\Di=\epsilon^{\;\prime}\Di\Ji$, $\Ji\gamma=\epsilon^{\;\prime\prime}\gamma\Ji$, where the $\epsilon$-sign table depends on the dimension $n$ modulo 8 of the spectral triple.
If both spectral triples obey Connes' $\epsilon$-sign table, it is seen that their product , defined in the straightforward way above, does not necessarily obey this $\epsilon$-sign table. In this note, we propose an alternative definition of the product real structure such that the $\epsilon$-sign table is also satisfied by the product.
\end{abstract}
\vskip 5cm
\noindent PACS numbers : 11.15.-q, 02.40.-k ; \\
Keywords : Noncommutative Geometry, Real Spectral Triples
%
%
\newpage
\section{Introduction}
\setcounter{equation}{0}
A real spectral triple $\Trip=\{\Al,\Hi,\Di,\Ji(,\gamma)\}$ \cite{Con1}, is given by a pre-$C^*$-algebra $\Al$
with a faithful *-representation by bounded operators ${\cal B}(\Hi)$ on a Hilbert space $\Hi$ : \(\pi:\Al\rightarrow{\cal B}(\Hi):a\rightarrow\pi(a)\).
A self-adjoint Dirac operator $\Di$ with compact resolvent acts on this Hilbert space such that \([\Di,\pi(a)]\in{\cal B}(\Hi)\).
The dimension of the spectral triple is given by the integer $n$ such that the operator $|\Di|^{-n}$ defined on $\Hi\setminus{\cal K}er(\Di)$ is an infinitesimal of first order. This means that the eigenvalues $\mu_\alpha$ of the compact positive operator $|\Di|^{-n}$ arranged in decreasing order $\mu_0\geq\mu_1\geq\cdots$ behave asymptotically as 
\(\mu_\alpha=O(\alpha^{-1})\) and \(\sigma_n:=\sum_{\alpha<N}\mu_\alpha=O(\log N)\), where the sum includes the multiplicities of the eigenvalues. The coefficient of $\log N$ is, by definition \cite{Con},\cite{Var}, the noncommutative integral of $|\Di|^{-n}$ written as \(\NCI |\Di|^{-n}\). The real structure $\Ji$ is an antilinear isometry in $\Hi$. It is further assumed that $\pi^o(a)=\Ji\pi(a)^\dagger\Ji^{-1}$ provides a representation of the opposite algebra $\Al^o$ commuting with $\pi(b),\forall b\in\Al$, so that $\Hi$ is endowed with an $\Al$-bimodule structure. The Dirac operator is then assumed to be a first-order operator on this bimodule which entails that \([[\Di,\pi(a)],\pi^o(b)]=0\). This real structure should further obey the relations :
\be\label{1} \Ji^2=\epsilon\;\Id\;;\;\Ji\Di=\epsilon^{\;\prime}\Di\Ji\;,\ee
where the {\it epsilons} are sign factors $\pm 1$. Finally, when the dimension is even, there is a grading operator $\gamma$, i.e. $\gamma^\dagger=\gamma$ and $\gamma^2=\Id$, such that the representation $\pi(a)$ of $a\in\Al$ is even, \(\pi(a)\gamma-\gamma\pi(a)=0\), and the Dirac operator is odd, $\Di\gamma+\gamma\Di=0$. The real structure $\Ji$ and the grading $\gamma$ obey also a relation of the type :
\be\label{2} \Ji\gamma=\epsilon^{\;\prime\prime}\gamma\Ji\;,\ee
where again $\epsilon^{\;\prime\prime}$ is a sign factor $\pm 1$.\\
In the typical commutative example, the algebra $\Al$ consists of the smooth functions on a compact Riemannian spin-manifold $M$ and the Hilbert space $\Hi$ is made of the square integrable spinors $\Psi(x)$ on $M$. The representation of $\Al$ on $\Hi$ is just the multiplication of $\Psi(x)$ by functions $f(x)$. $\Di$ is then the usual (massless!) Dirac operator $\Di=-\im\gamma^\mu\nabla_\mu$, where the hermitian $\gamma$-matrices obey $\gamma^\mu\gamma^\nu+\gamma^\nu\gamma^\mu=+2\delta^{\mu\nu}$ and where $\nabla_\mu$ is the covariant derivative acting on spinor fields. The dimension $n$ is then the usual dimension of the manifold $M$ and \(\NCI |\Di|^{-n}\) is proportionnal to the volume of $M$. The real structure $\Ji$ generalizes the (Euclidean!) charge conjugation operation 
\(\Con \Psi(x)=C\Psi^*(x)\), where \(C{\gamma^\mu}^*C^{-1}=-\gamma^\mu\). When the dimension of $M$ is even, $n=2k$, the grading is given by the chirality matrix $\gamma_{2k+1}=(\im)^{k}\gamma^1\gamma^2\cdots\gamma^{2k}$, anti-commuting with all $\gamma^\mu$ matrices and with the Dirac operator $\Di$.\\
Connes \cite{Con2} showed that the {\it epsilons}, defined in \ref{1},\ref{2}, may be determined by the dimension n, modulo 8, due to Bott periodicity of the real Clifford algebras. They are given in table \ref{T1}.\\ 
\noindent
In applications to particle models, one usually takes the product of a commutative spectral triple as above, where $M$ is a configuration space or a Riemannian (Wick rotated) space-time, with a 0-dimensional genuinely noncommutative spectral triple describing the internal structure of the particles. Since nowadays almost any dimension is on the model-building market, it seems useful to examine in general the definition of the product of two real spectral triples $\Trip_1=\{\Al_1,\Hi_1,\Di_1,\Ji_1,\gamma_1\}$ and $\Trip_2=\{\Al_2,\Hi_2,\Di_2,\Ji_2(,\gamma_2)\}$. With the first necessarily even, it is defined \cite{Con2} as $\Trip=\{\Al,\Hi,\Di,\Ji(,\gamma)\}$ where $\Al=\Al_1\otimes\Al_2$, $\Hi=\Hi_1\otimes\Hi_2$, $\Ji=\Ji_1\otimes\Ji_2$ and $\Di=\Di_1\otimes\Id_2+\gamma_1\otimes\Di_2$. This implies that $\Di^2=(\Di_1)^2\otimes\Id_2+\Id_1\otimes(\Di_2)^2$ and the dimensions add : $n=n_1+n_2$. When $\Trip_2$ is also even, the total grading is given by $\gamma=\gamma_1\otimes\gamma_2$. In order that this recipe should work, with $\Trip$ obeying the $\epsilon$-sign table, the following conditions should be met :
\bea
\Ji^2=\epsilon\;\Id\;,&\hbox{with}\;\epsilon=\epsilon_1\;\epsilon_2\;,\label{Eerste}\\
\Ji\Di=\epsilon^{\;\prime}\Di\Ji\;,&\hbox{with}\;\epsilon^{\;\prime}=
\epsilon^{\;\prime}_1=\epsilon^{\;\prime\prime}_1\epsilon^{\;\prime}_2\;,\label{Tweede}\\
\hbox{and,}&\hbox{when both are even,}\nonumber\\
\Ji\gamma=\epsilon^{\;\prime\prime}\gamma\Ji\;,&\hbox{with}\;
\epsilon^{\;\prime\prime}=\epsilon^{\;\prime\prime}_1\;\epsilon^{\;\prime\prime}_2\;.\label{Derde}\eea
\newpage\noindent
A prompt examination of the $\epsilon$-sign table shows that, in the even-even case, condition (\ref{Derde}) is satisfied\footnote{There is a more sophisticated proof \cite{Kraj} available, invoking the image under $\pi_\Di$ of the  Hochschild $n_k$-cycles, instead of the mere inspection of the $\epsilon$-sign table.}. However it is readily seen from table \ref{T2} and \ref{T3} that at least one of the conditions (\ref{Eerste}),(\ref{Tweede}) is violated, in the even-even case, when $n_1\in\{6,2\}$ and, in the odd case, when the total dimension $n\in\{5,1\}$. In the even-even case, we could transform the Dirac operator $\Di$ with the unitary operator 
\(\U=\frac{1}{2}\left(\Id_1\otimes\Id_2+\gamma_1\otimes\Id_2+\Id_1\otimes\gamma_2-\gamma_1\otimes\gamma_2\right)\),
so that \(\Di^{\;\prime}=\U\Di\U^\dagger=\Di_1\otimes\gamma_2+\Id_1\otimes\Di_2\). The condition (\ref{Tweede}) is then obeyed in cases $n_1\in\{6,2\}$ and $n_2\in\{4,0\}$ with $\Ji=\Ji_1\otimes\Ji_2$ but, when $n_1\in\{6,2\}$ and $n_2\in\{6,2\}$, it is not. In the even-odd case there is even no $\gamma_2$ available. Also a modification of the individual Dirac operators with the unitary operator \(\V=\left(\Id_1+\im\gamma_1\right)/\sqrt 2\), changing $\Di_1$ into $\im\gamma_1\Di_1$ will not help much in matching condition (\ref{Eerste}). It appears thus that it is $\Ji$ that has to be modified.
\section{The modified product real structure}
\setcounter{equation}{0}
The clue in changing the real structure of the even spectral triple lies in the property of \(\widetilde{\Ji_1}=\Ji_1\gamma_1\) which is such that 
\be\label{Premier}(\widetilde{\Ji_1})^2=\widetilde{\epsilon_1}\;\Id_1\;\ee
where $\tilde\epsilon_1=\epsilon_1\;\epsilon^{\;\prime\prime}_1$ remains unchanged for $n_1\in\{4,0\}$ but changes sign for $n_1\in\{6,2\}$. Furthermore \be\label{Segond}\widetilde{\Ji_1}\Di_1=\tilde\epsilon^{\;\prime}_1\Di_1\;,\ee with \(\tilde\epsilon^{\;\prime}_1=-\epsilon^{\;\prime}_1\). Finally, since $\epsilon^{\;\prime\prime}_1$ does not change, condition (\ref{Derde}) remains satisfied.
The even-odd cases with $n\in\{5,1\}$ are readily cured defining the product real structure as \(\Ji=\widetilde{\Ji_1}\otimes\Ji_2=\Ji_1\gamma_1\otimes\Ji_2\) as table \ref{T4} shows.
In the even-even case, when $n_1\in\{6,2\}$, the $\epsilon_1^{\;\prime}$ should not change sign since, for all even $n$, 
$\epsilon^{\;\prime}=\epsilon^{\;\prime}_1=+1$. To recover the original $+$ sign we multiply by $\gamma=\gamma_1\otimes\gamma_2$ so that the real structure of the product reads \(\Ji=\left(\Ji_1\gamma_1\otimes\Ji_2\right)\left(\gamma_1\otimes\gamma_2\right)=\Ji_1\otimes\Ji_2\gamma_2\).
The \(n_1\in\{6,2\}\) cases will then, with this real structure, obey Connes' $\epsilon$-sign table as the table \ref{T5} shows.
\section{Conclusions}
In this short note, we have redefined, by elementary algebraic techniques, the real structure of the product of two real spectral triples such that Connes' $\epsilon$-sign table remains valid for the product if it holds for each factor. This is achieved taking as real structure $\Ji$ given by :
\begin{itemize}
\item $\Ji=\Ji_1\gamma_1\otimes\Ji_2$ when $n_1+n_2=n\in\{5,1\}$,
\item $\Ji=\Ji_1\otimes\Ji_2\gamma_2$ when $n_1\in\{6,2\}$ and $n_2$ even
\item $\Ji=\Ji_1\otimes\Ji_2$ in all other cases.
\end{itemize}
\vspace{1cm}
{\bf Acknowledgements}\\
We thank prof.Joseph C.V\'{a}rilly of UCR,San Jos\'{e},Costa Rica, for discussions concerning the $\epsilon$-sign table before and during the $X^{th}$ J.A.Swieca Summer School, section : Particles and Fields, \'{A}guas de Lind\'{o}ia, february,1999.

\newpage
\begin{table}\caption{The $\epsilon$-sign table}\label{T1}\begin{center}
\begin{tabular}{|l||l|l|l|l|l|l|l|l|} \hline
n=                         & 0 & 1 & 2 & 3 & 4 & 5 & 6 & 7 \\ \hline\hline
$\epsilon$                 & + & + & - & - & - & - & + & + \\ \hline
$\epsilon^{\;\prime}$      & + & - & + & + & + & - & + & + \\ \hline
$\epsilon^{\;\prime\prime}$& + & $\star$ & - & $\star$ & + & $\star$ & - & $\star$ \\ \hline
\end{tabular}\end{center}\end{table}
\newpage
\begin{table}\caption{The even-even case}\label{T2}\begin{center}
\begin{tabular}{|l|l|l||c|c|c|c|c||c|c|c|c|c|c|} \hline
$n_1$ & $n_2$ & $n$ & $\epsilon_1$ & $\epsilon_2$ & $\epsilon_1\;\epsilon_2$ & $\epsilon$ & ?                          
& $\epsilon_1^{\;\prime}$ & $\epsilon_1^{\;\prime\prime}$ & $\epsilon_2^{\;\prime}$ & 
$\epsilon_1^{\;\prime\prime}\;\epsilon_2^{\;\prime}$ & $\epsilon^{\;\prime}$ & ? \\ \hline\hline
6 & 6 & 4 & + & + & + & - & N & + & - & + & - & + & N \\ \hline
6 & 4 & 2 & + & - & - & - & Y & + & - & + & - & + & N \\ \hline
6 & 2 & 0 & + & - & - & + & N & + & - & + & - & + & N \\ \hline
6 & 0 & 6 & + & + & + & + & Y & + & - & + & - & + & N \\ \hline
4 & 6 & 2 & - & + & - & - & Y & + & + & + & + & + & Y \\ \hline
4 & 4 & 0 & - & - & + & + & Y & + & + & + & + & + & Y \\ \hline
4 & 2 & 6 & - & - & + & + & Y & + & + & + & + & + & Y \\ \hline
4 & 0 & 4 & - & + & - & - & Y & + & + & + & + & + & Y \\ \hline
2 & 6 & 0 & - & + & - & + & N & + & - & + & - & + & N \\ \hline
2 & 4 & 6 & - & - & + & + & Y & + & - & + & - & + & N \\ \hline
2 & 2 & 4 & - & - & + & - & N & + & - & + & - & + & N \\ \hline
2 & 0 & 2 & - & + & - & - & Y & + & - & + & - & + & N \\ \hline
0 & 6 & 6 & + & + & + & + & Y & + & + & + & + & + & Y \\ \hline
0 & 4 & 4 & + & - & - & - & Y & + & + & + & + & + & Y \\ \hline
0 & 2 & 2 & + & - & - & - & Y & + & + & + & + & + & Y \\ \hline
0 & 0 & 0 & + & + & + & + & Y & + & + & + & + & + & Y \\ \hline
\end{tabular}\end{center}\end{table}
\newpage
\begin{table}\caption{The even-odd case}\label{T3}\begin{center}
\begin{tabular}{|l|l|l||c|c|c|c|c||c|c|c|c|c|c|} \hline
$n_1$ & $n_2$ & $n$ & $\epsilon_1$ & $\epsilon_2$ & $\epsilon_1\;\epsilon_2$ & $\epsilon$ & ?                          
& $\epsilon_1^{\;\prime}$ & $\epsilon_1^{\;\prime\prime}$ & $\epsilon_2^{\;\prime}$ & 
$\epsilon_1^{\;\prime\prime}\;\epsilon_2^{\;\prime}$ & $\epsilon^{\;\prime}$ & ? \\ \hline\hline
6 & 7 & 5 & + & + & + & - & N & + & - & + & - & - & N \\ \hline
6 & 5 & 3 & + & - & - & - & Y & + & - & - & + & + & Y \\ \hline
6 & 3 & 1 & + & - & - & + & N & + & - & + & - & - & N \\ \hline
6 & 1 & 7 & + & + & + & + & Y & + & - & - & + & + & Y \\ \hline
4 & 7 & 3 & - & + & - & - & Y & + & + & + & - & + & Y \\ \hline
4 & 5 & 1 & - & - & + & + & Y & + & + & - & + & - & N \\ \hline
4 & 3 & 7 & - & - & + & + & Y & + & + & + & - & + & Y \\ \hline
4 & 1 & 5 & - & + & - & - & Y & + & + & - & + & - & N \\ \hline
2 & 7 & 1 & - & + & - & + & N & + & - & + & - & - & N \\ \hline
2 & 5 & 7 & - & - & + & + & Y & + & - & - & + & + & Y \\ \hline
2 & 3 & 5 & - & - & + & - & N & + & - & + & - & - & N \\ \hline
2 & 1 & 3 & - & + & - & - & Y & + & - & - & + & + & Y \\ \hline
0 & 7 & 7 & + & + & + & + & Y & + & + & + & + & + & Y \\ \hline
0 & 5 & 5 & + & - & - & - & Y & + & + & - & - & - & N \\ \hline
0 & 3 & 3 & + & - & - & - & Y & + & + & + & + & + & Y \\ \hline
0 & 1 & 1 & + & + & + & + & Y & + & + & - & - & - & N \\ \hline
\end{tabular}\end{center}\end{table}
\newpage
\begin{table}\caption{The cured even-odd case}\label{T4}\begin{center}
\begin{tabular}{|l|l|l||c|c|c|c|c||c|c|c|c|c|c|} \hline
$n_1$ & $n_2$ & $n$ & $\tilde\epsilon_1$ & $\epsilon_2$ & $\epsilon_1\;\epsilon_2$ & $\epsilon$ & ?                          
& $\tilde\epsilon_1^{\;\prime}$ & $\epsilon_1^{\;\prime\prime}$ & $\epsilon_2^{\;\prime}$ & 
$\epsilon_1^{\;\prime\prime}\;\epsilon_2^{\;\prime}$ & $\epsilon^{\;\prime}$ & ? \\ \hline\hline
6 & 7 & 5 & - & + & - & - & Y & - & - & + & - & - & Y \\ \hline
6 & 3 & 1 & - & - & + & + & Y & - & - & + & - & - & Y \\ \hline
4 & 5 & 1 & - & - & + & + & Y & - & + & - & + & - & Y \\ \hline
4 & 1 & 5 & - & + & - & - & Y & - & + & - & + & - & Y \\ \hline
2 & 7 & 1 & + & + & + & + & Y & - & - & + & - & - & Y \\ \hline
2 & 3 & 5 & + & - & - & - & Y & - & - & + & - & - & Y \\ \hline
0 & 5 & 5 & + & - & - & - & Y & - & + & - & - & - & Y \\ \hline
0 & 1 & 1 & + & + & + & + & Y & - & + & - & - & - & Y \\ \hline
\end{tabular}\end{center}\end{table}
\newpage
\begin{table}\caption{The cured even-even cases}\label{T5}\begin{center}
\begin{tabular}{|l|l|l||c|c|c|c|c||c|c|c|c|c|c|} \hline
$n_1$ & $n_2$ & $n$ & $\epsilon_1$ & $\tilde\epsilon_2$ & $\epsilon_1\;\tilde\epsilon_2$ & $\epsilon$ & ?                          
& $\epsilon_1^{\;\prime}$ & $\epsilon_1^{\;\prime\prime}$ & $\tilde\epsilon_2^{\;\prime}$ & 
$\epsilon_1^{\;\prime\prime}\;\tilde\epsilon_2^{\;\prime}$ & $\epsilon^{\;\prime}$ & ? \\ \hline\hline
6 & 6 & 4 & + & - & - & - & Y & + & - & - & + & + & Y \\ \hline
6 & 4 & 2 & + & - & - & - & Y & + & - & - & + & + & Y \\ \hline
6 & 2 & 0 & + & + & + & + & Y & + & - & - & + & + & Y \\ \hline
6 & 0 & 6 & + & + & + & + & Y & + & - & - & + & + & Y \\ \hline
2 & 6 & 0 & - & - & + & + & Y & + & - & - & + & + & Y \\ \hline
2 & 4 & 6 & - & - & + & + & Y & + & - & - & + & + & Y \\ \hline
2 & 2 & 4 & - & + & - & - & Y & + & - & - & + & + & Y \\ \hline
2 & 0 & 2 & - & + & - & - & Y & + & - & - & + & + & Y \\ \hline
\end{tabular}\end{center}\end{table}
%
\end{document}